\documentclass[aip,apl,reprint,amsmath,amssymb,nofootinbib]{revtex4-2}

\usepackage{graphicx}
\usepackage{siunitx}
\usepackage{bm}

\begin{document}

\title{Spatially resolved elastic strain and lattice rotation at threading
dislocations in HgCdTe/CdZnTe epilayers by dark-field X-ray microscopy}

\author{C.~Yildirim}
\email{can.yildirim@esrf.fr}
\affiliation{European Synchrotron Radiation Facility, 71 Avenue des Martyrs, 38000 Grenoble, France}

\author{A.~Benhadjira}
\affiliation{European Synchrotron Radiation Facility, 71 Avenue des Martyrs, 38000 Grenoble, France}

\author{P.~Ballet}
\affiliation{Universit\'e Grenoble Alpes, CEA, LETI, 38000 Grenoble, France}

\author{T.~N.~Tran Caliste}
\affiliation{European Synchrotron Radiation Facility, 71 Avenue des Martyrs, 38000 Grenoble, France}

\author{J.~Baruchel}
\affiliation{European Synchrotron Radiation Facility, 71 Avenue des Martyrs, 38000 Grenoble, France}

\author{C.~Detlefs}
\affiliation{European Synchrotron Radiation Facility, 71 Avenue des Martyrs, 38000 Grenoble, France}

\author{D.~Brellier}
\affiliation{Universit\'e Grenoble Alpes, CEA, LETI, 38000 Grenoble, France}

\author{M.~P.~Kabukcuoglu}
\affiliation{Laboratory for Applications of Synchrotron Radiation , Karlsruhe Institute of Technology, 76131, Karlsruhe, Germany}

\date{\today}

\begin{abstract}
Threading dislocations (TDs) propagating from a Cd$_{1-y}$Zn$_{y}$Te (CZT)
substrate into a liquid-phase-epitaxy Hg$_{1-x}$Cd$_{x}$Te (MCT) epilayer set the
minority-carrier lifetime and dark-current floor of mid-wave infrared
focal-plane arrays, yet at device-grade densities their local strain fields have
been accessible only through topography, which conflates lattice tilt and elastic
strain. We apply dark-field X-ray microscopy in reflection geometry to a
\SI{7}{\micro\metre}-thick (111) MCT/CZT epilayer. Shallow Bragg angle and absorption makes the signal layer
dominated while the numerical aperture of the objective keeps the layer and
substrate rocking curves convolved, so weak-beam images on either side of the
rocking curve and their difference image the correlated defects in a single
frame: dot-like substrate TDs and the elongated, in-plane island features they
nucleate in the layer. Kernel average misorientation resolves each TD as a
\SI{7}{\micro\metre} signature, the layer thickness, alongside axial strain lobes
of $\pm(4$ to $5)\times10^{-5}$.
\end{abstract}

\maketitle

Hg$_{1-x}$Cd$_{x}$Te (MCT) is the benchmark photovoltaic material for
high-performance infrared detectors across the short-, mid-, and long-wave
bands, owing to its composition-tunable direct band gap, large absorption
coefficient, and compatibility with large-format focal-plane
arrays.\cite{rogalski2005,lei2015} As detector technologies push toward smaller
pixels and lower dark current, the way atomic-scale defects perturb the
surrounding lattice becomes increasingly consequential. Threading dislocations
(TDs) that emerge from the Cd$_{1-y}$Zn$_{y}$Te (CZT) substrate and propagate
through the epilayer act as Shockley--Read--Hall recombination centres and diode
shunt paths, and they set the minority-carrier lifetime and the dark-current
floor of the finished array.\cite{johnson1992,shin1992} Each dislocation also
carries a long-range elastic strain field that locally modifies the band
structure and the electrical behaviour of the material around
it.\cite{johnson1992} Controlling TDs in device-grade material therefore requires
non-destructive tools that resolve not only their density but also their
crystallographic character and their local strain fields, over areas large enough
to be representative.

At the TD densities of state-of-the-art liquid-phase-epitaxy (LPE)
MCT/CZT,\cite{brellier2014} $10^{3}$ to $10^{4}~\si{\per\square\centi\metre}$,
most established probes are limited. Etch-pit counting is destructive and needs
material-specific chemistries.\cite{everson1995,hahnert1990} Atomic force
microscopy resolves individual TD intersections with the surface but, with a
field of view of at most ${\sim}100\times100~\si{\micro\metre\squared}$, makes
statistical sampling impractical, while optical interferometry trades resolution
for area, and neither reaches the buried interface or the strain field around a
core.\cite{fourreau2016} Transmission electron microscopy maps strain directly
over only tens of unit cells, its specimen preparation can introduce dislocations
into this material, and the TD density is too low for representative counts per
foil.\cite{fourreau2016} X-ray Bragg diffraction imaging and its quantitative
form, rocking-curve imaging, address the mapping problem
non-destructively,\cite{lubbert2000,tranthi2017} but a scalar rocking curve
measures only the effective misorientation
\begin{equation}
\Delta\theta \;=\; \delta\theta \;+\; \frac{\Delta d}{d}\,\tan\theta_\mathrm{B},
\label{eq:misorientation}
\end{equation}
where $\delta\theta$ is the rigid rotation of the diffracting planes and
$\Delta d/d$ the local axial strain. A single rocking curve cannot separate the
two, so the elastic strain field of an individual TD, the quantity that matters
for device performance, remains convolved with the lattice rotations.

Dark-field X-ray microscopy (DFXM)\cite{simons2015,poulsen2017,yildirim2020mrs}
lifts this limitation. An objective lens in the diffracted beam forms a magnified
image of the illuminated volume and acts as a pinhole in reciprocal space,
linking the image to a restricted, known region of it; rocking the sample angle
$\phi$ at fixed $2\theta$ maps lattice rotation, while scanning $2\theta$ maps
axial strain.\cite{poulsen2017,poulsen2018rsm} Because DFXM can share the
reflection-geometry diffraction condition of topography, the two can be run in
succession on one instrument to build a multiscale picture of the same
volume.\cite{simons2019,jakobsen2019} DFXM has resolved dislocation structures in
bulk metals,\cite{jakobsen2019,yildirim2023} mapped the strain and rotation
fields of individual misfit dislocations in epitaxial oxides,\cite{simons2019}
and followed dislocation dynamics at high temperature.\cite{dresselhaus2021} It
has been applied to an operable MCT detector,\cite{yildirim2021situ} but not to
the strain field of individual dislocations in the II--VI MCT/CZT system. Here we
apply DFXM in reflection geometry to the same LPE MCT/CZT sample we characterised
previously by near-field X-ray topography,\cite{yildirim2021} and report the
first maps of the elastic strain and lattice-rotation fields of individual dislocation structures in
this system.

\begin{figure}
\includegraphics[width=\columnwidth]{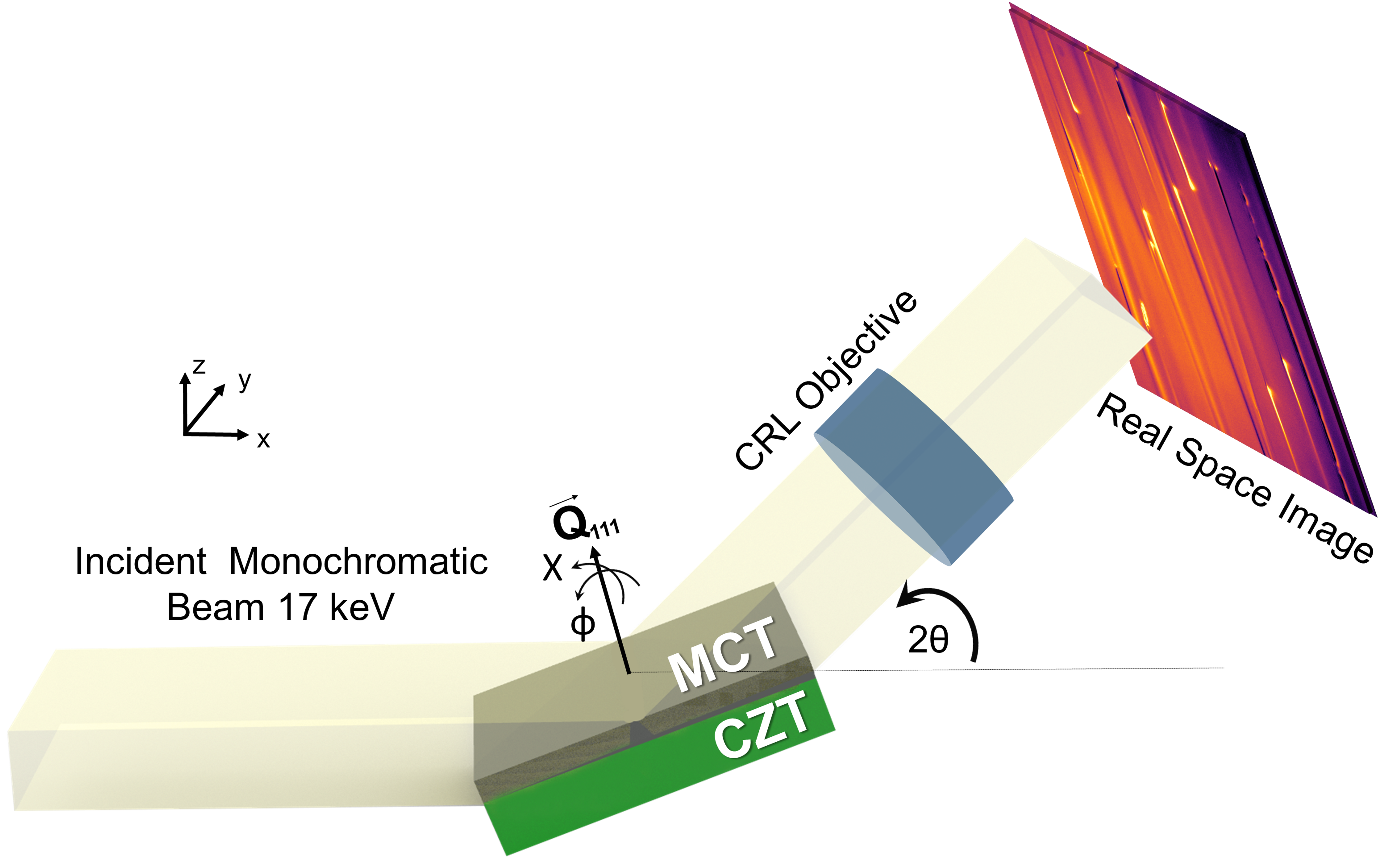}
\caption{DFXM in reflection geometry on the MCT/CZT epilayer. A $\phi$ scan at fixed $2\theta$ maps the lattice tilt, a $2\theta$
scan the axial strain.}
\label{fig:setup}
\end{figure}

Measurements were performed at
beamline ID06-HXM (now ID03 \cite{isern2025}) of the European Synchrotron Radiation Facility
.\cite{kutsal2019} Monochromatic \SI{17}{\kilo\electronvolt} X-rays were
selected by a Si(111)  monochromator 
and pre-focused with two-dimensional Be compound refractive lenses (CRLs) in the
optics hutch. The beam footprint was about
$400\times400~\si{\micro\metre\squared}$ (horizontal $\times$ vertical), and all
measurements were carried out in reflection geometry on the 111 reflection with $2\theta=\ang{11.21}$ (Fig.~\ref{fig:setup}). For sample alignment and
near-field topography, a FReLoN CCD indirect detector with
\SI{0.62}{\micro\metre} effective pixel size, placed \SI{50}{\milli\metre}
downstream of the sample. For DFXM, a CRL
objective comprising 88 Be parabolic lenslets ($R=\SI{50}{\micro\metre}$) was inserted
\SI{281}{\milli\metre} downstream, giving an X-ray magnification
$M_\mathrm{x}=17.5$. The diffracted beam was converted to visible light by a
scintillator and recorded by a PCO.edge sCMOS camera ($2160\times2560$ pixels)
through a visible-light microscope about \SI{5}{\metre} from the sample.
Switching the visible objective between $10\times$ and $2\times$ gave total
magnifications $M_t\approx179$ and $36$, corresponding to effective sample-plane
pixel sizes of about \SI{42}{\nano\metre} and \SI{210}{\nano\metre}. Because the
measurements are in reflection geometry, images are stretched along the beam
direction by $1/\sin\theta_\mathrm{B}$ relative to true sample coordinates.

The sample is the LPE MCT/CZT heterostructure of
Ref.~\onlinecite{yildirim2021}: a \SI{7}{\micro\metre}-thick
Hg$_{0.71}$Cd$_{0.29}$Te epilayer grown by LPE\cite{pelliciari1994} on a
double-side-polished (111)-oriented CZT substrate, with
TD density in the $10^{3}$ to $10^{4}\,\si{\per\square\centi\metre}$ range. The Zn
fraction $y$ was tuned to minimise the lattice mismatch and to avoid the
cross-hatch of misfit dislocations.\cite{brellier2014,tobin1995} Axial-strain
$\varepsilon_{33}=\Delta d/d$ and tilt $\delta\theta$ maps were reconstructed
from the first moments of the $(\phi,2\theta)$ scans using the standard DFXM
decomposition.\cite{poulsen2017,darfix} The lattice-tilt field is shown with a
two-dimensional key in which hue encodes the tilt direction $(\chi,\phi)$ and
lightness its magnitude.\cite{darfix}

Before presenting the results, an important point to note is which crystal the DFXM maps describe. Two properties of this diffraction
condition fix what the images contain. In reflection geometry a beam diffracted
at depth $z$ traverses $2z/\sin\theta_\mathrm{B}$ of MCT, so the diffracted
intensity falls by $1/e$ every \SI{1.2}{\micro\metre} of depth and the signal is
layer dominated. The decay is exponential and not a cut-off: the contribution per
unit depth at the buried interface is still $0.3\%$ of that at the free surface,
and because the CZT substrate is a high-quality crystal with large integrated
reflectivity per unit volume, the first one to two micrometres below the
interface still register. Layer and substrate are separated in rocking angle by
$\Delta\phi=(\Delta d/d)\tan\theta_\mathrm{B}$, which for the
$\Delta d/d=1.2\pm0.2\times10^{-4}$ measured on this sample\cite{yildirim2021} is a small fraction of the
rocking-curve width, while the reciprocal-space acceptance of the
objective\cite{poulsen2017} is more than an order of magnitude larger than that
separation. Both contributions are therefore transferred to every image and their
rocking curves are convolved rather than resolved, two overlapping peaks of very
unequal weight. The maps below are consequently maps of the epilayer that retain
a weak but coherent imprint of the substrate immediately beneath it, and it is
this that makes the correlated defects of Ref.~\onlinecite{yildirim2021} visible
in a single frame.

\begin{figure*}
\includegraphics[width=\textwidth]{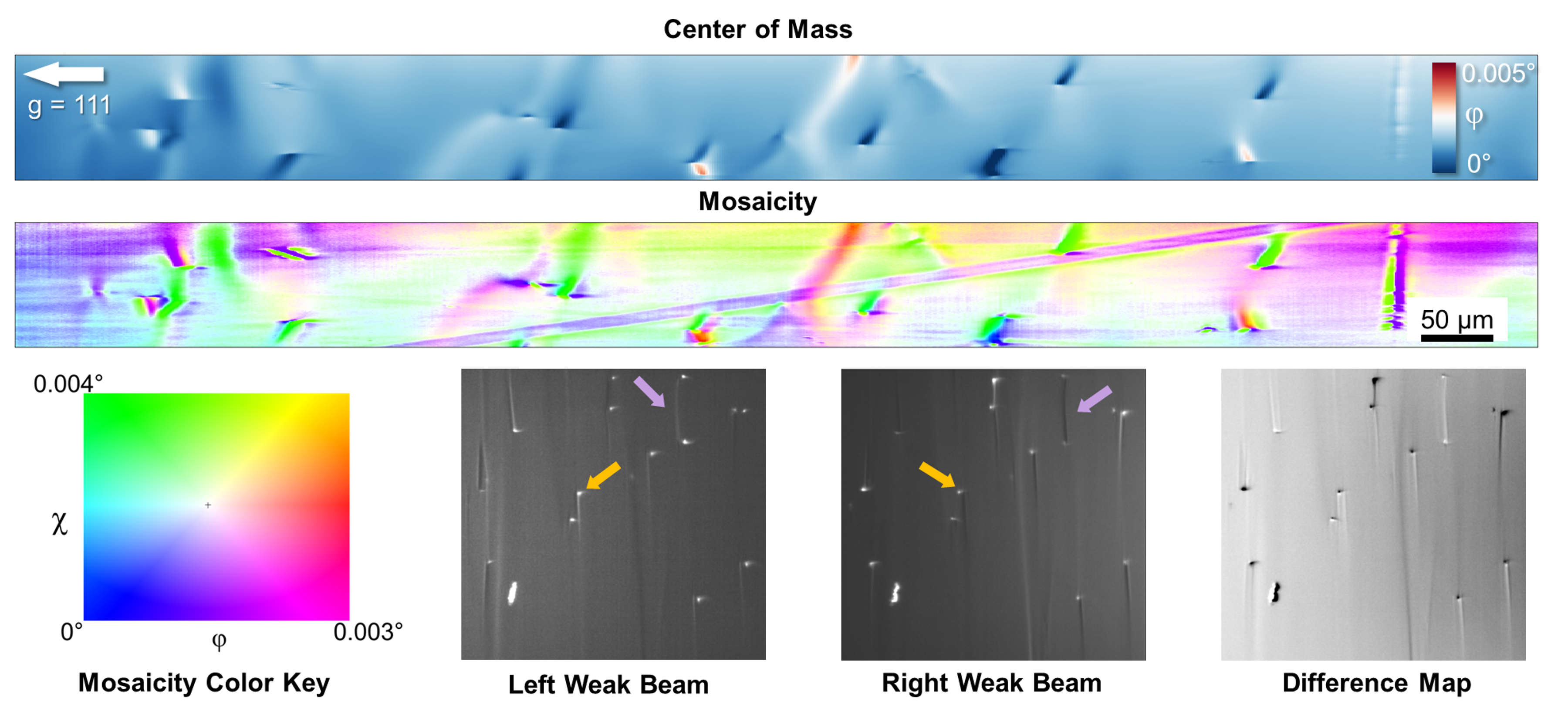}
\caption{Lattice rotation around individual threading dislocations imaged with
the $10\times$ objective on the (111) reflection. Top: centre-of-mass map of the
rocking angle $\phi_\mathrm{COM}$ Middle: two-dimensional mosaicity map of the same
field, hue encoding the tilt direction $(\chi,\phi)$ and lightness its magnitude.
Dislocations appear as orientation singularities where the hue wraps around the
core, and a long dislocation line crosses the field diagonally. Bottom: mosaicity colour key, raw images at the low- and
high-$\theta$ weak-beam settings, and their difference. Weak beam and difference map were not corrected for geometrical projection for better visualisation.}
\label{fig:mosaicity}
\end{figure*}

Figure~\ref{fig:mosaicity} maps the lattice rotation around individual TDs with
the $10\times$ objective. The COM map of the rocking angle,
$\phi_\mathrm{COM}$ (top), shows a gently varying matrix punctuated by sharp
dipolar features at the cores, while the two-dimensional mosaicity map (middle)
renders the tilt direction and magnitude in colour. Individual dislocations
appear as orientation singularities where the hue wraps around the core, and a
long dislocation line crosses the field diagonally. The rotation field extends
over tens of micrometres around each defect, in clear contrast to the elastic
strain, which we show later in the manuscript stays within a few micrometres of the cores. The
dot-like and wavy line-like features that were unresolved in the millimetre-scale
topographs of Ref.~\onlinecite{yildirim2021}.

The lower row of
Fig.~\ref{fig:mosaicity} shows raw images recorded on the low- and
high-$\theta$ (corresponding to $\phi$ rotation) flanks of the rocking curve together with their difference. Two
families of contrast are present. Compact, dot-like features (yellow arrows)
invert their brightness between the two settings and appear with opposite sign in
the difference map; since the substrate peak lies to higher angle, the
high-$\theta$ setting weights its contribution most, and we identify these with
the TDs emerging from the CZT substrate, the dot-like features of
Ref.~\onlinecite{yildirim2021}. Elongated, wavy features (purple arrows) keep
their sign; these are the island-like features mentioned in the previous work, and they
correspond to defects that nucleate in the epilayer around the emerging substrate
TDs. Because the difference map carries both families in one frame, it images the
correlated defects of Ref.~\onlinecite{yildirim2021} directly rather than by
comparing separately reconstructed maps.

A systematic observation is that the elongated features lie in the layer plane
and run perpendicular to the projected direction of $\bm{g}$. The two tilt
directions are not equivalent with respect to such features. DFXM measures the changes around the local scattering vector, whose two components are the in-plane
gradients of the out-of-plane displacement $u_g=\bm{u}\cdot\hat{\bm{g}}$: the
rocking angle $\phi$ responds to $\partial u_g/\partial x$, the gradient along
the beam-projected direction, and $\chi$ to $\partial u_g/\partial y$, the
gradient across it. A defect elongated along $y$ has a displacement field that
varies steeply across itself and hardly at all along its length, so its
signature falls almost entirely in $\phi$, while a segment elongated along $x$
does the reverse and appears in $\chi$. Any map built from a $\phi$ scan
therefore renders preferentially the segments running perpendicular to the
projected $\bm{g}$. 

The gently varying, wavy background of the $\phi_\mathrm{COM}$ map follows from
the convolution described above. The centroid of two overlapping peaks
separated by $\Delta\phi$ lies between them at a position set by their
relative weight, so anything that modulates the substrate contribution shifts
$\phi_\mathrm{COM}$ without any lattice rotation being involved. That weight is
exponentially sensitive to the local epilayer thickness through
$\exp[-2t/(\Lambda\sin\theta_\mathrm{B})]$, and it also follows the local
diffracting power of the substrate, which is highest where its own dislocations
break extinction. The wavy contrast is therefore not a property of the epilayer
alone, and it is seen most directly in the $10\times$ rocking-curve movie given
in the supplementary material.

\begin{figure*}
\includegraphics[width=\textwidth]{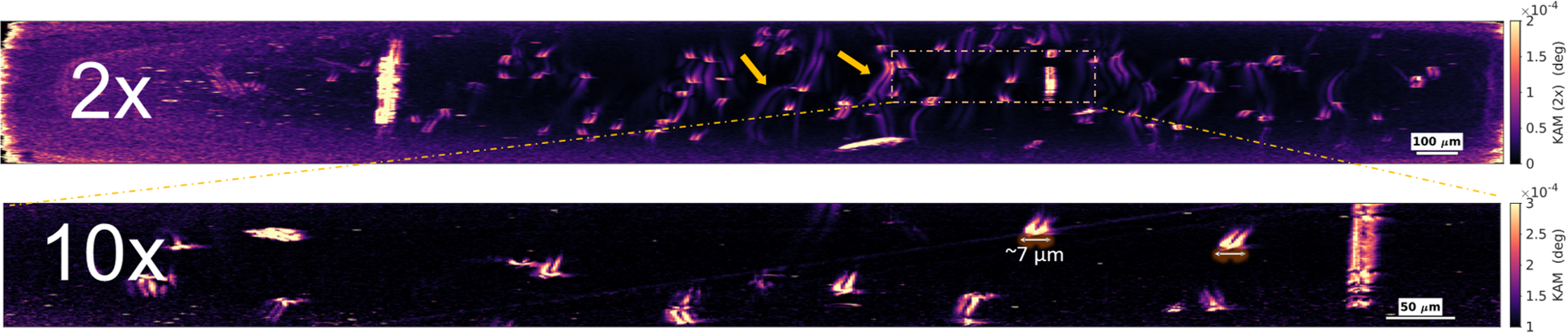}
\caption{Multiscale kernel average misorientation (KAM) of the threading
dislocations. Top: KAM
map at $2\times$. The dashed box marks the region re-imaged at higher
magnification. Bottom: KAM map of that region at $10\times$, resolving individual
dislocation structures' misorientation gradient. The KAM
signature of each dislocation extends about \SI{7}{\micro\metre} along the
beam-projected direction. }
\label{fig:kam}
\end{figure*}

Figure~\ref{fig:kam} maps the kernel average misorientation (KAM), the local
magnitude of the lattice-orientation gradient, at two magnifications. These maps show similar (but inverted) contrasts to the amplitude (integrated-intensity) map shown in Supplementary material. In the
wide-field $2\times$ map the dislocations stand out as bright filaments against
the low-KAM matrix, and the $10\times$ map resolves each one as a pair of
comet-shaped prongs. The two prongs of a pair are separated by about $7~\mu$m
along the beam-projected direction, the same separation at $2\times$ and
$10\times$ and equal to the epilayer thickness; for a near-normal line of length
$t$ imaged in symmetric reflection at $\theta_\mathrm{B}\approx8.8^{\circ}$ the
projected separation is $t\cos\theta_\mathrm{B}\approx t$. Each pair is therefore
a single dislocation imaged across the depth of the film, with one prong at its
emergence from the buried MCT/CZT interface and the other at its termination at
the free surface, joined by a short zigzag that follows the inclined line. In the
orthogonal in-plane direction, some of the same dislocations run continuously over hundreds
of micrometres, bending across the field; these are the bright lines that
connected neighbouring dislocations in the near-field topographs of
Ref.~\onlinecite{yildirim2021}, now resolved as individual curved segments.
Atomic force microscopy of this sample resolves, at the surface, a localised
distortion of the growth steps with a footprint of a few micrometres where a
dislocation reaches the free surface (Ref.~\onlinecite{yildirim2021}, Fig.~4), the
surface signature of the same prong that DFXM images through the layer.

\begin{figure*}
\includegraphics[width=\textwidth]{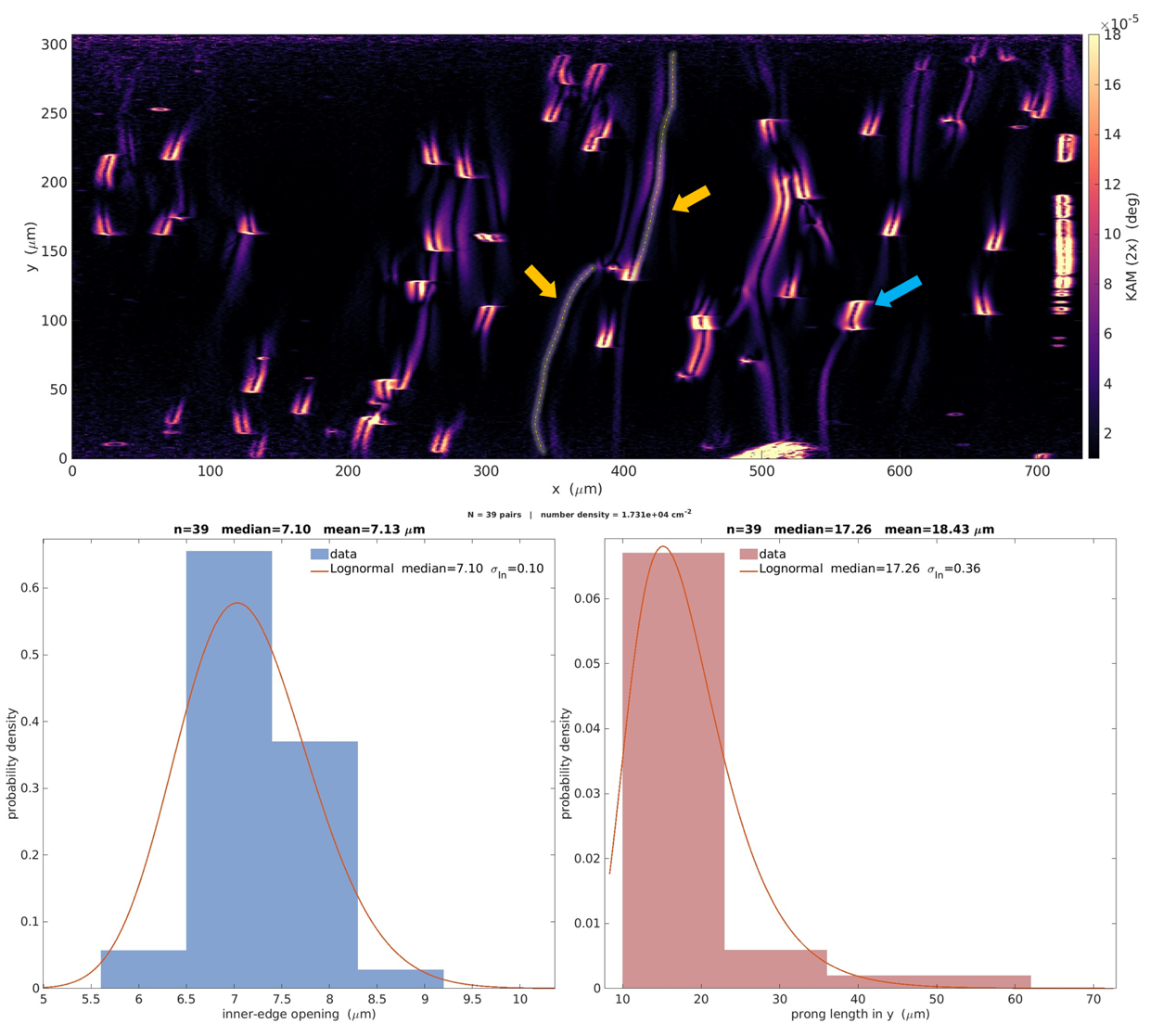}
\caption{Statistics of the dislocation signatures in the 2$\times$-objective KAM map,
$730\times300~\si{\micro\metre\squared}$ field. Each dislocation appears as a pair
of bright prongs (blue arrow), the projected emergence at the 
interface and termination at the free surface, while individual lines run and
bend laterally across the field over tens of micrometres (orange arrows). A
total of $N=39$ dislocation associated pairs were analysed, giving a
threading-dislocation number density of
$1.73\times10^{4}~\si{\per\square\centi\metre}$, of order the
$10^{3}$ to $10^{4}~\si{\per\square\centi\metre}$ density of the substrate. Lower
left: inner-edge opening, the gap between the two prongs of each pair, median
\SI{7.10}{\micro\metre}, with a narrow lognormal fit ($\sigma_{\ln}=0.10$)
essentially fixed at the \SI{7}{\micro\metre} epilayer thickness; the values below
\SI{7}{\micro\metre} correspond to lines terminating at the free surface within
the layer. Lower right: prong length along $y$, median \SI{17.26}{\micro\metre},
with a broader fit ($\sigma_{\ln}=0.36$) reflecting the variable in-plane extent
of the segments.}
\label{fig:stats}
\end{figure*}

We quantified these dislocation structures over the wide-field KAM map (Fig.~\ref{fig:stats}).
Across $39$ dislocation-associated pairs the inner-edge opening, the gap between the two
prongs of each pair, is narrowly distributed with a median of $7.10~\mu$m and a
lognormal width $\sigma_{\ln}=0.10$, fixed at the $7~\mu$m epilayer thickness. The
prong length along the layer normal is broader, with a median of $17.26~\mu$m and
$\sigma_{\ln}=0.36$. The contrast between the two distributions is the central
result of the statistics: the prong separation is set by a geometric constant, the
layer thickness, and is therefore narrow, while the prong length follows the
variable inclination and in-plane excursion of each line and is therefore broad.
The number density of the pairs, $1.7\times10^{4}~\mathrm{cm}^{-2}$, is of the
order of the substrate dislocation density and agrees with the near-field and
etch-pit values of Ref.~\onlinecite{yildirim2021}, confirming that the resolved
features are substrate threading dislocations propagating through the epilayer and
not an artefact of the imaging. 

\begin{figure*}
\includegraphics[width=\textwidth]{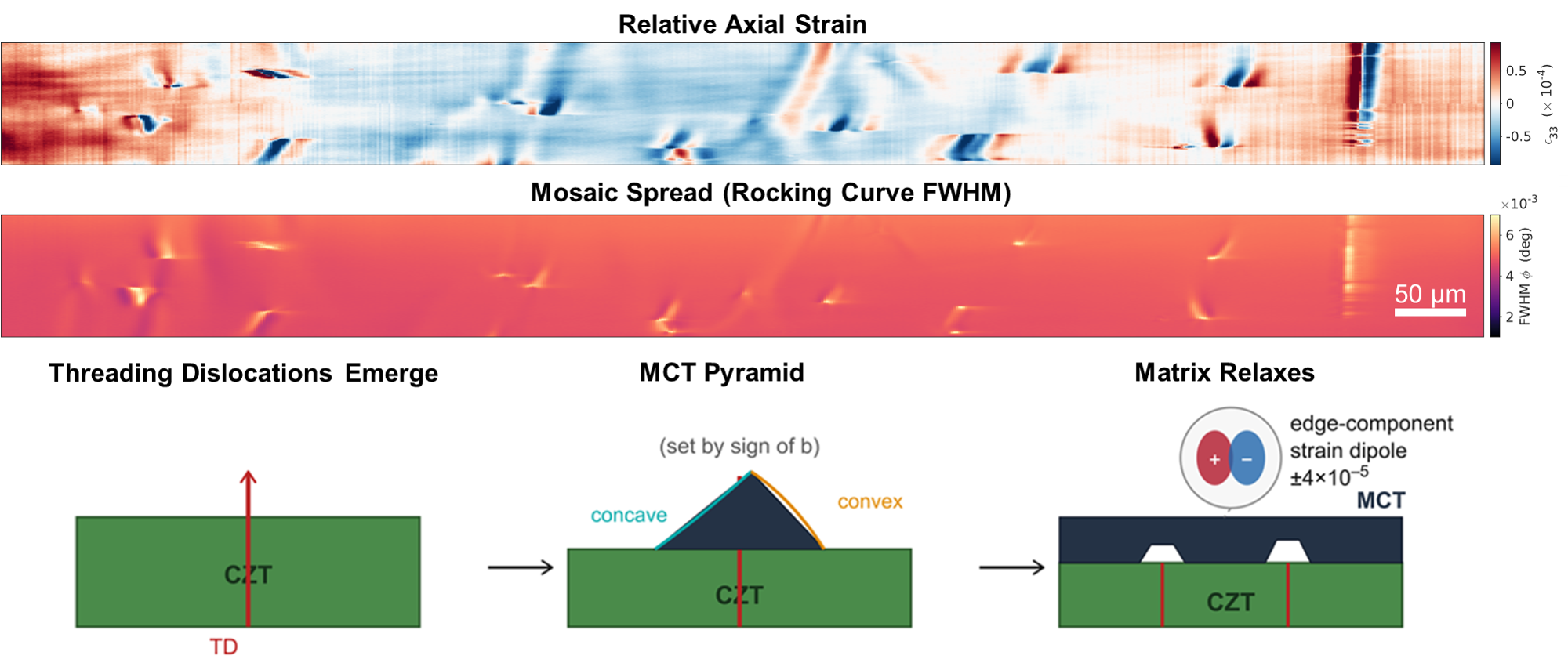}
\caption{ (a)~Axial-strain map $\varepsilon_{33}$ (b)~Rocking-curve
full width at half maximum over the same field (c)~Dislocation-templated growth: a substrate
TD emerges and offers a favoured growth ledge (1); an MCT pyramid nucleates with
$\Delta a/a\approx1\times10^{-4}$ and an asymmetric, Burgers-vector-dependent
accommodation, concave on one side and convex on the other (2); the pyramids
coalesce into the mature matrix, leaving the edge-component strain lobe of
$\pm(4$ to $5)\times10^{-5}$ resolved in (a)}
\label{fig:strain}
\end{figure*}

To examine the associated strain fields, we turn to the DFXM measurements. Figure~\ref{fig:strain}(a,b) shows the axial strain, $\varepsilon_{33}$, and the rocking-curve full width at half maximum over a wide field of view. The two maps reveal distinct aspects of the defect structure. The axial strain remains close to zero across most of the matrix, within approximately $\pm0.5\times10^{-4}$, and is concentrated into compact features within a few micrometres of the dislocation cores. In contrast, the rocking-curve width highlights the cores as bright streaks marking regions of enhanced lattice distortion.

The strain contrast does not resemble the simple two-lobe dipole expected for a straight edge dislocation viewed end-on. Instead, the axial strain alternates in sign around each core, producing tensile–compressive–tensile and, in some cases, tensile–compressive–tensile–compressive sequences. This multipolar pattern can be attributed to the mixed character and curved trajectory of the TDs. Because the DFXM image integrates the strain field through the full $7~\mu$m epilayer thickness, the projection of the dislocation strain onto the 111 diffraction vector changes with depth as the local line direction and edge–screw character vary. The dipolar strain fields associated with the interface-emergence and surface-termination segments are therefore laterally offset and partially overlap in projection, giving rise to the observed alternating lobes. A similar superposition effect has been reported for misfit dislocations in epitaxial BiFeO$_3$.\cite{simons2019}

The measured strain amplitudes are consistent with $\tfrac{1}{2}\langle110\rangle$ glide dislocations. For MCT with $a=6.47~\text{\AA}$ and $|\mathbf{b}|=a/\sqrt{2}=4.58~\text{\AA}$, the edge component produces an axial strain of order $7\times10^{-5}$ at a distance of $1~\mu$m from the line, using $u\sim b/(2\pi r)$.\cite{hirth1992,kaganer2005} This agrees with the resolved values of approximately $\pm(4$–$5)\times10^{-5}$ after projection onto the diffraction vector. The orientation and polarity of the strain lobes are analogous to the contrast used to identify threading dislocations by electron channelling in GaN,\cite{kamaladasa2011} where the lobe geometry depends on the Burgers vector and diffraction condition. Consistent with this interpretation, dislocation-contrast simulations for the three $\mathbf{g}$-visible $\tfrac{1}{2}\langle110\rangle$ variants reproduce the observed behaviour, and the partition between edge-like and screw-like character is independently recovered from the Nye tensor components $\alpha_{31}$, $\alpha_{32}$ and $\alpha_{33}$ calculated from the measured $\varepsilon_{33}$, $\phi$ and $\chi$ maps. We support these observations with dislocation-type simulations shown in supplementary information.


A clearer picture now emerges of how TDs originating in the CZT substrate influence epilayer growth. Defects generated during growth remain correlated with the TD network over distances of tens of micrometres, while the dislocations evolve in character as they propagate toward the surface. These maps give quantitative support to the
dislocation-templated growth picture of Ref.~\onlinecite{yildirim2021}. A substrate TD emerging at the growth front provides
an energetically favoured ledge for MCT attachment\cite{frank1949,quere1998} and
nucleates a local growth pyramid in which the MCT templates onto the substrate
lattice parameter, departing from it by $\Delta a/a=1.2\times10^{-4}$ as growth
advances.\cite{yildirim2021} Because the depositing MCT has a larger lattice
parameter than the substrate, the accommodation around the emerging TD is
asymmetric, concave on one side of the line and convex on the other, with the
favoured side set by the sign of the Burgers vector. The localised strain lobe at
each core is the real-space image of this asymmetry, and the progressive build-up
of mismatch between the substrate-templated material around each TD and the
relaxed matrix is what drives the dislocations to bend over and run in-plane,
acquiring the misfit-segment character that relieves the increasing mismatch as
the layer thickens. This is the origin of the elongated "island" (wavy-line) features that the weak-beam images separate from the substrate dots, and the correlation between
the two families in the difference map is the direct image of it. The long-range
tilt domains are the collective elastic accommodation of these extended, curving
networks. Finally, small, localised axial strain coexisting with large,
long-range tilt is the fingerprint of mixed TDs with a non-negligible screw
component along the $<111>$ growth direction: a pure edge dislocation threading
111 would dominate the axial-strain map, while a pure screw would be silent
in $\varepsilon_{33}$ but contribute strongly to the rotation field. This is
consistent with the spiral-growth model proposed
earlier,\cite{yildirim2021,frank1949} in which screw components emerging at the
surface set the handedness of growth spirals.

In a narrow-gap material such as mid-wave MCT the band
gap is sensitive to strain through the deformation potentials, so the
$\pm(4$ to $5)\times10^{-5}$ axial strain and lattice distortions more than 0.001$\circ$ around TDs translate into local fluctuations of the band gap. These fluctuations may act as recombination and trapping
sites.\cite{johnson1992,shin1992} Because the fields reach micrometres to tens of micrometres, the electrically active footprint of a TD is set by its strain field, not by the nanometer-scale dislocation core alone; and because the tilt domains span 50 to \SI{100}{\micro\metre}, comparable to a focal-plane pixel, a single dislocation network can produce pixel-to-pixel non-uniformity in responsivity and dark current. Mapping where the strain is, how it is partitioned between edge-like and screw-like accommodation, and how far it spreads points to the growth levers that reduce it: substrate quality, the Zn fraction that sets the lattice match,\cite{tobin1995,brellier2014} surface preparation, and growth rate. Paired with automated, learning-based classification of dislocation families,\cite{abdou2025} this becomes a routine route to strain-aware defect engineering in II--VI infrared-detector materials.

In conclusion, using DFXM, we show that the elastic strain and lattice-rotation fields of individual TDs in an LPE-grown Hg$_{1-x}$Cd$_{x}$Te epilayer. Absorption makes the signal layer dominated while the aperture of the objective keeps the layer and substrate rocking curves convolved, so weak-beam images on either side of the rocking curve and their difference image the correlated defects in a single frame, separating the dot-like substrate TDs from the elongated, in-plane wavy-line features they nucleate in the layer and accounting for the wavy background of the tilt maps. The strain to tilt decoupling exposes localised, alternating-sign strain lobes within about \SI{5}{\micro\metre} of each core and a long-range rotation field an order of magnitude larger in angle that carries the screw-component signature. KAM maps show that each TD results in correlated layer defects that travels the full \SI{7}{\micro\metre} layer while running and bending laterally
over tens of micrometres.

\begin{acknowledgments}
The authors thank the ESRF for beamtime at ID06-HXM and the ID06 staff for their support during the experiment. We acknowledge J.~Merlin and R.~Templier for complementary characterisation of the sample. CY acknowledge the financial support from the ERC Starting Grant "D-REX" nr 10116911.
\end{acknowledgments}

\section*{Supplementary material}
See the supplementary material for the near-field topographic overview of the same region, integrated-intensity maps of the $10\times$  scans along with a rocking-curve movie of the same scan, the Nye dislocation density tensor analysis, and the dislocation-contrast simulations for pure edge, pure screw and mixed character.

\section*{Data availability}
The data that support the findings of this study are available from the corresponding author upon reasonable request.

\bibliographystyle{aipnum4-2}
\bibliography{references}

\end{document}


\title{SUPPLEMENTARY MATERIAL}
\maketitle

\section{Near field Topography}
In order to have an overview of the sample, we collected rocking curve imaging on the near field, covering a much larger area to show the statistical relevance of the dislocation structures. Similar to the study, Yildirim et al \cite{yildirim2021}, this map is corrected for curvature of the film and the divergence of the incident beam. The central part of the map is where the zoomed in 10x and 2x studies are shown in the main manuscript.
\begin{figure}[H]
    
    \includegraphics[width=1\linewidth]{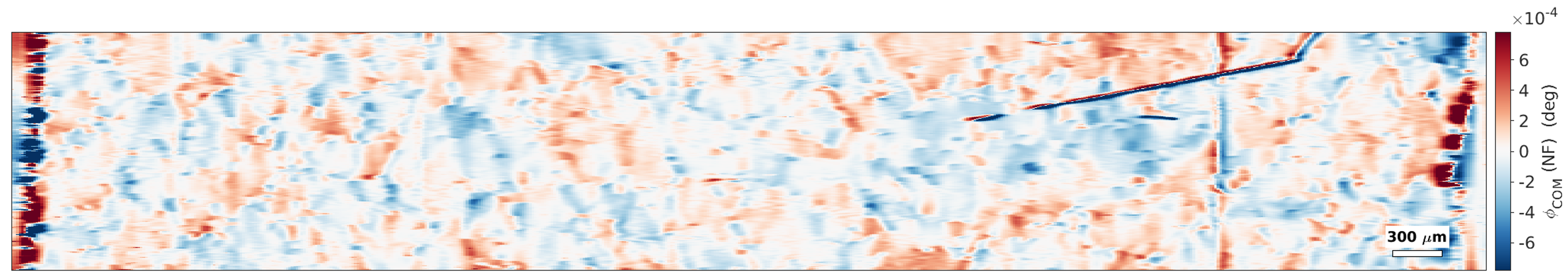}
    \caption{Rocking curve center of mass map. Images were collected on a near field detector that was positioned 50mm downstream the sample.}
    
    \end{figure}

\section{Integrated Intensity Maps}
Integrated intensity maps are calculated for each pixel by fitting a gaussian function as a function of sample tilt. These maps show the sum of the diffracted intensity over the rocking curve per each pixel, and are analogous to the KAM maps shown in the main text.

\begin{figure}[H]
    
    \centering
    \includegraphics[width=0.5\linewidth]{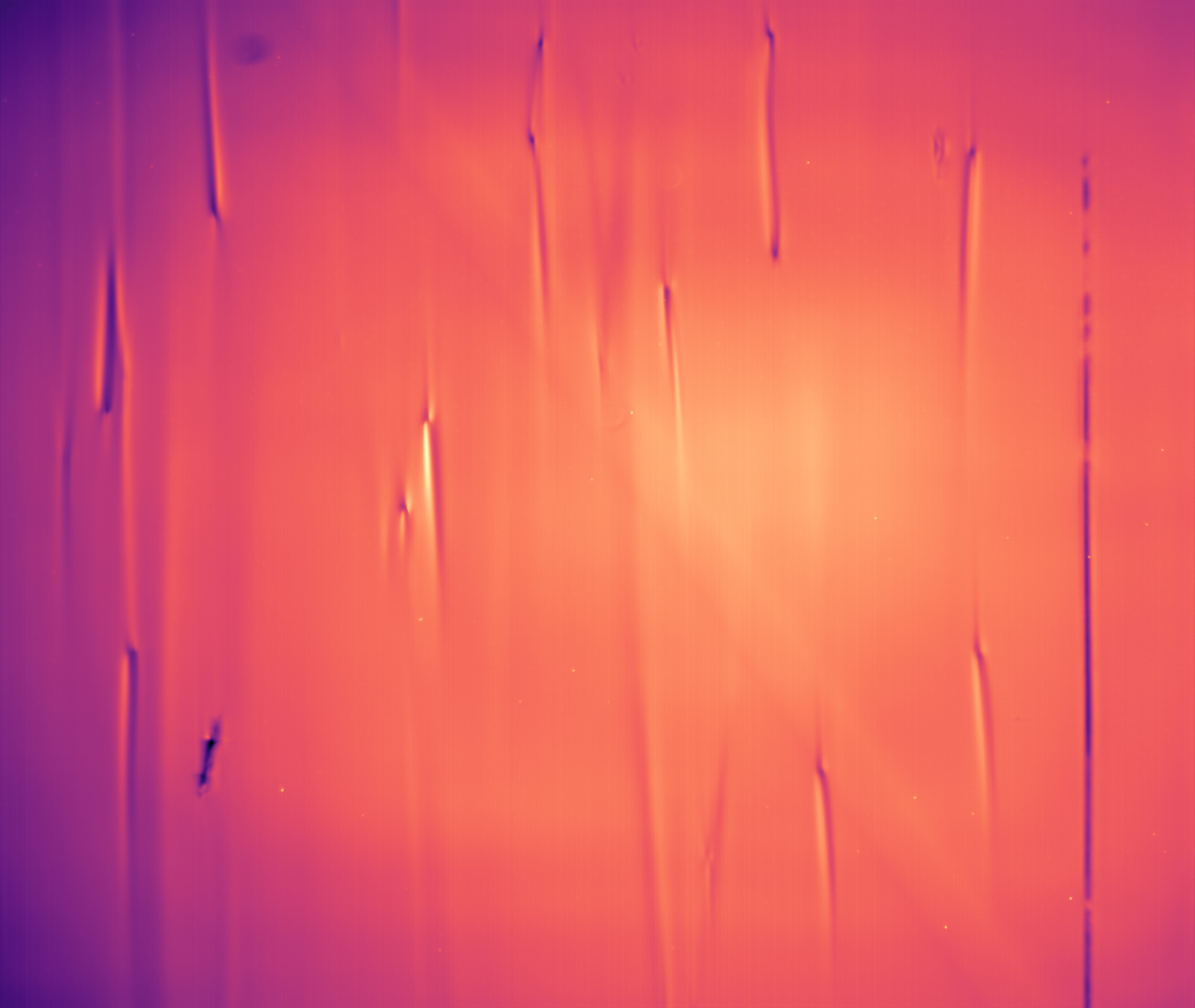}
    \caption{Integrated intensity map of the rocking curve shown in the main text colelcted at 10$\times$ objective. The field of view is the same as Figure 2 of the manuscript. Image aspect ratio is not corrected for the geometrical stretching.}
    
    \end{figure}

\section{Nye tensor analysis and consistency of the measured strain amplitudes}

From the DFXM mosaicity and strain map on a given reflection, one can derive three of the the dislocation density tensor components. Here we show the experimentally derived Nye tensor components
$\alpha_{31}$, $\alpha_{32}$ and $\alpha_{33}$ obtained from the DFXM
strain and orientation maps according to

\begin{align}
\alpha_{31} &= -\frac{\partial \varepsilon_{33}}{\partial x_2},\\
\alpha_{32} &= \phantom{-}\frac{\partial \varepsilon_{33}}{\partial x_1},\\
\alpha_{33} &= -\left(
\frac{\partial \phi}{\partial x_2}
+
\frac{\partial \chi}{\partial x_1}
\right),
\end{align}

following the formalism of Simons \emph{et al.}\cite{simons2019}.
The components $\alpha_{31}$ and $\alpha_{32}$ are sensitive to
edge-character dislocations, whereas $\alpha_{33}$ reflects screw
character. The measured values are significantly smaller than those
reported for threading dislocations imaged directly in BiFeO$_3$,
where $\alpha_{33}$ values of order
$2\times10^{-4}\,\mathrm{nm^{-1}}$ were associated with individual
dislocations occupying a single pixel.\cite{simons2019}

This reduction is expected because the present measurements probe
the long-range strain field of an inclined and curved dislocation
segment after projection onto the diffraction vector and integration
through approximately $7~\mu$m of material. Oppositely signed strain
lobes generated by different parts of the dislocation line therefore
partially overlap in projection, reducing the apparent strain
gradients and the resulting Nye tensor amplitudes.

The measured strain amplitudes are nevertheless quantitatively
consistent with elasticity theory. For a
$\frac{1}{2}\langle110\rangle$ perfect dislocation in MCT
($a=6.47$~\AA, $b=a/\sqrt{2}=4.58$~\AA), the far-field displacement
gradient scales approximately as

\begin{equation}
\varepsilon \sim \frac{b}{2\pi r}.
\end{equation}

At a distance of $r=1~\mu$m from the dislocation line this yields

\begin{equation}
\varepsilon \approx
\frac{4.58\times10^{-10}}
     {2\pi\times10^{-6}}
\simeq 7\times10^{-5},
\end{equation}

which is of the same order as the experimentally observed
$\varepsilon_{33}$ amplitudes of approximately
$\times10^{-4}$.
The agreement supports the interpretation that the observed
alternating strain lobes originate from the elastic field of a
single curved dislocation whose projected strain field is modified by
depth integration and overlap of contributions from different
segments of the line.

\begin{figure}[H]
        \includegraphics[width=\linewidth]{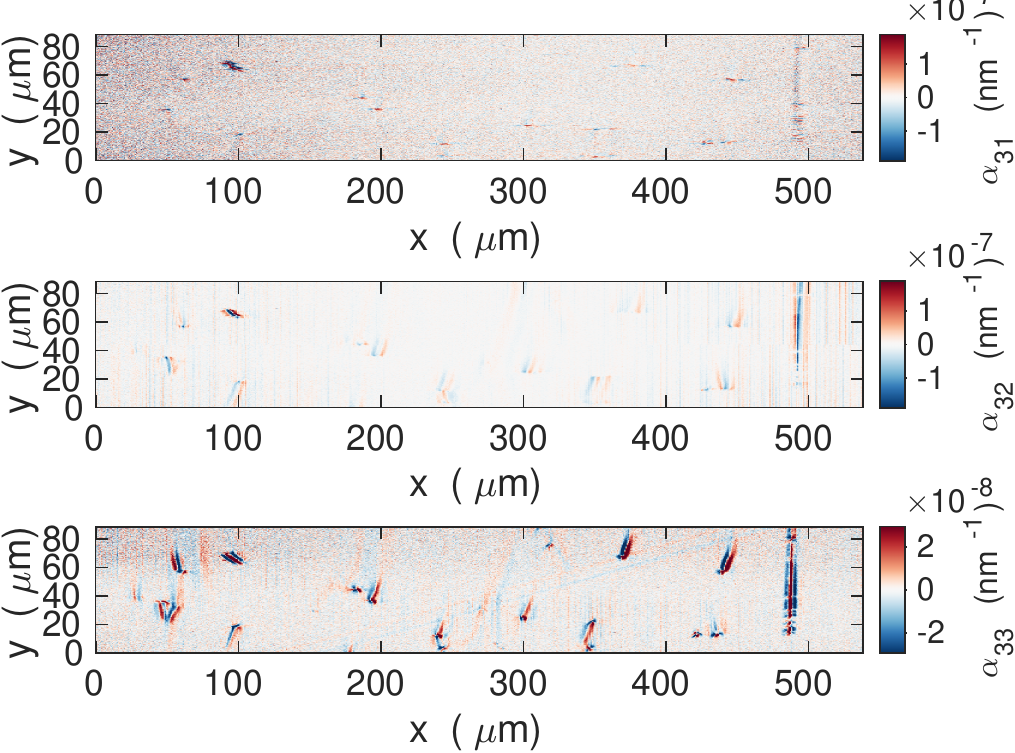}
        \caption{Spatially resolved Nye tensor components derived from DFXM strain and orientation maps. Maps of the resolvable Nye dislocation tensor components (a) ($\alpha_{31}$), (b) ($\alpha_{32}$), and (c) ($\alpha_{33}$), calculated from the measured axial strain ($\varepsilon_{33}$) and lattice rotations ($\phi$) and ($\chi$) according to ($\alpha_{31}=-\partial\varepsilon_{33}/\partial x_2$), ($\alpha_{32}=\partial\varepsilon_{33}/\partial x_1$), and ($\alpha_{33}=-(\partial\phi/\partial x_2+\partial\chi/\partial x_1)$). Following the interpretation of Simons \emph{et al.}, ($\alpha_{31}$) and ($\alpha_{32}$) are associated with edge-dislocation character, whereas ($\alpha_{33}$) reflects screw-dislocation character. Positive and negative values correspond to opposite signs of the edge component or opposite screw chirality. The elongated alternating features observed in ($\alpha_{31}$) and ($\alpha_{32}$) coincide with the strain lobes in the ($\varepsilon_{33}$) maps and indicate spatially varying edge-character dislocation density. The comparatively weaker ($\alpha_{33}$) signal is consistent with a predominantly edge-like defect configuration. All components are shown in units of nm($^{-1}$); a (2$\times$2) median filter was applied for visualization.}
\end{figure}

\section{Simulations}

A dislocation-contrast simulation was performed for an FCC thin film in (111) Bragg-reflection geometry. Three dislocations were arranged in a triangular configuration, with Burgers vectors chosen from the visible $\langle 110\rangle$ variants according to the $\mathbf{g}\cdot\mathbf{b}$ condition: $[110]$, $[101]$, and $[011]$. Pure edge, pure screw, and mixed edge--screw characters were compared.

\begin{equation}
u_g(\mathbf{r})=\mathbf{u}(\mathbf{r})\cdot\hat{\mathbf{g}}
\end{equation}

\begin{equation}
\varepsilon_g(\mathbf{r}) \simeq \frac{\partial u_g}{\partial z}
\end{equation}

\begin{equation}
\Delta\phi(\mathbf{r}) \simeq -\tan(\theta_B)\,\varepsilon_g(\mathbf{r})
\end{equation}

The finite-thickness Bragg reflectivity was approximated using a two-beam dynamical diffraction  \cite{Carlsen2022a,Carlsen2022b},

\begin{equation}
R(\phi)=\left|\frac{E_h(\phi)}{E_0^{\mathrm{inc}}}\right|^2 ,
\end{equation}

where \(E_0\) and \(E_h\) are the incident and diffracted wave amplitudes. The simulated reflected intensity was obtained by sampling the rocking curve at the locally shifted Bragg angle,

\begin{equation}
I(\mathbf{r},\phi)
=
I_0(\mathbf{r})
R\!\left[\phi-\Delta\phi(\mathbf{r})\right].
\end{equation}

\begin{figure}
    \centering
    \includegraphics[width=\linewidth]{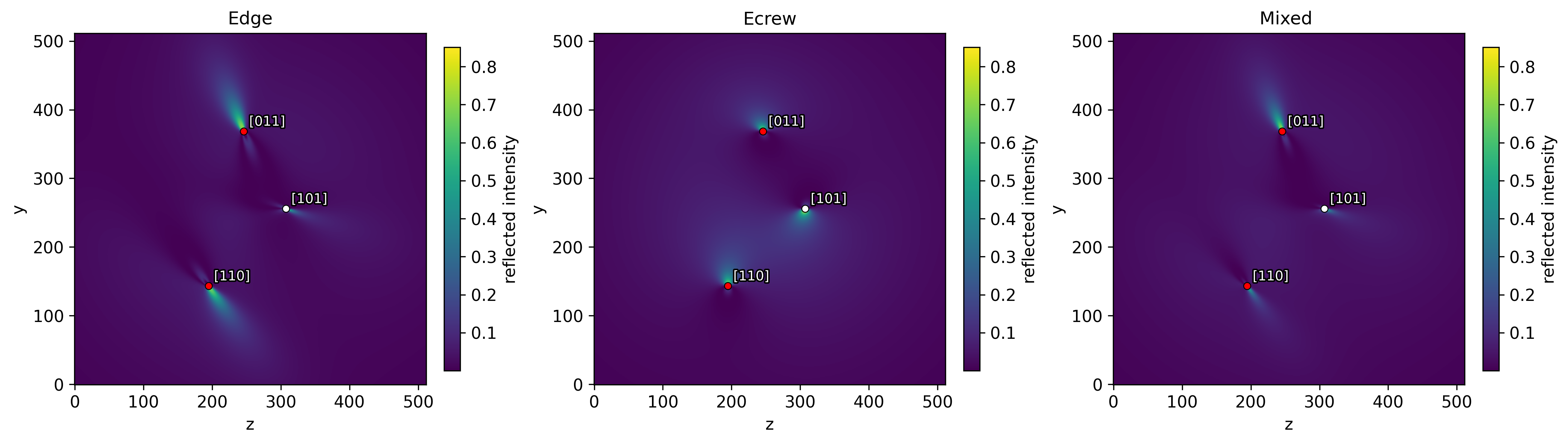}
    \caption{Simulated 111 Bragg-reflection contrast for an FCC thin film with a triangular arrangement of dislocations, with the three possible $\langle 110\rangle$ Burgers-vector variants selected according to the $\mathbf{g}\cdot\mathbf{b}$ visibility condition. The dislocations are labeled by their Burgers vectors, $[110]$, $[101]$, and $[011]$. The panels compare pure edge, pure screw, and mixed edge--screw character at a fixed weak beam condition at $\phi = -1.0 \times 10^{-4}\,\mathrm{rad}$.}
    \label{fig:sim}
\end{figure}

This model is intended to reproduce the qualitative dependence of the contrast on Burgers vector and dislocation character rather than to provide a fully quantitative prediction of the measured intensity. Images were evaluated at $\phi=-1.0\times10^{-4}\,\mathrm{rad}$ as shown in Fig.~\ref{fig:sim}.

\bibliographystyle{abbrv}
\bibliography{references}